# Curie temperature versus hole concentration in field-effect structures of $Ga_{1-x}Mn_xAs$

Y. Nishitani,[1] D. Chiba,[2,1] M. Endo,[1] M. Sawicki,[3,1] F. Matsukura,[1,2] T. Dietl,[1,2,3,4] and H. Ohno[1,2]

[1]*Laboratory for Nanoelectronics and Spintronics, Research Institute of Electrical Communication, Tohoku University, Katahira 2-1-1, Aoba-ku, Sendai 980-8577, Japan*
[2]*Semiconductor Spintronics Project, Exploratory Research for Advanced Technology, Japan Science and Technology Agency, Sanban-cho 5, Chiyoda-ku, Tokyo 102-0075, Japan*
[3]*Institute of Physics, Polish Academy of Sciences, al. Lotników 32/46, PL-02-668 Warszawa, Poland*
[4]*Institute of Theoretical Physics, University of Warsaw, ul. Hoża 69, PL-00-681 Warszawa, Poland*

The Curie temperature $T_C$ is investigated as a function of the hole concentration $p$ in thin films of ferromagnetic semiconductor (Ga,Mn)As. The magnetic properties are probed by transport measurements and $p$ is varied by the application of an external electric field in a field-effect transistor configuration. It is found that $T_C \sim p^\gamma$, where the exponent $\gamma = 0.19 \pm 0.02$ over a wide range of Mn compositions and channel thicknesses. The magnitude of $\gamma$ is reproduced by a $p$-$d$ Zener model taking into account nonuniform hole distribution along the growth direction, determined by interface states and the applied gate electric fields.

## I. INTRODUCTION

The use of III-V ferromagnetic semiconductor (Ga,Mn)As has made it possible to demonstrate a number of new principles of spintronic device operations.[1-3] A large number of experimental and theoretical efforts have been devoted to raise the Curie temperature of (Ga,Mn)As, $T_C$ - the reported highest value of $T_C$ is now at a 185 K level,[4] a three-fold increase from the initial 60 K.[5] However, $T_C$ has not surpassed room temperature. Because the ferromagnetic coupling is mediated by holes in (Ga,Mn)As,[6] the hole concentration $p$ is one of the key parameters determining $T_C$. In most of previous experimental studies the hole concentration has been changed by growth conditions and/or post-growth low-temperature annealing,[7-10] which affect the density of double-donor defects, such as As antisites ($As_{Ga}$) (Ref. 7) and Mn interstitials ($Mn_{int}$) (Ref. 11) that compensate Ga-substitutional Mn acceptors, $Mn_{Ga}$. Because (Ga,Mn)As films have to be grown by low-temperature molecular-beam epitaxy (LT-MBE) to overcome the solubility limit,[5] a precise control of growth conditions is necessary,[12] which is not always easy.[13] Thus, in many experiments, low-temperature annealing has been employed to increase $p$ (Refs. 8-11) and hence $T_C$, according to the $p$-$d$ Zener model.[6] It is now known that the increase in $p$ results from diffusion of $Mn_{int}$ toward the surface, where they undergo oxidation.[14] In addition, the reduction in the density of $Mn_{int}$ increases the effective composition $x_{eff}$ of Mn participating in the ferromagnetic order because $Mn_{int}$ form antiferromagnetic pairs with $Mn_{Ga}$.[15] An extra hole doping by non-magnetic acceptors is also known to promote the formation of $Mn_{int}$.[16] Therefore, independent control of $p$ and $x_{eff}$ is difficult by changing growth parameters and/or annealing conditions. On the other hand, one of the most important characteristics of these semiconducting materials is the controllability of the carrier concentration by an external means, such as gate electric fields.[17-21] This approach can be applied to vary $T_C$ without changing the density of $Mn_{int}$ and, thus, $x_{eff}$.

In this paper, we report on the investigation of the relationship between $T_C$ and $p$ for (Ga,Mn)As films consisting a channel of field-effect transistor (FET) structures, in which the sheet hole concentration $p_{sheet}$ can be varied by applying a gate electric field.

According to the $p$–$d$ Zener model,[6] the magnitude of $T_C$ depends on the density of states and, thus, on the hole concentration. It is desirable to develop FET structures with large controllability of the hole concentration. We employ $Al_2O_3$ or $HfO_2$ formed by atomic layer deposition (ALD) as a gate insulator.[19] These oxides have dielectric constants $\kappa$ greater than 7 and high breakdown electric fields,[19-22] allowing for obtaining larger values of the product of capacitance $C$ and the gate voltage $V_G$ than those in the previous works on (In,Mn)As (Refs. 17 and 18) and (Ga,Mn)As (Ref. 23) with a spun-on gate insulator. To a first approximation $CV_G$ determines the modulation swing of $p_{sheet}$. Furthermore, larger changes in $p$ ($= p_{sheet}/t$) are achieved for thinner channels. Thus, we have fabricated structures with thin (Ga,Mn)As channels (thickness $t = 3.5 - 5$ nm).

## II. SAMPLE PREPARATION AND EXPERIMENTS

Twelve (Ga,Mn)As films with various nominal Mn composition $x$ ($= 0.052 - 0.20$) and $t$ ($= 3.5 - 5$ nm) have been grown by LT-MBE.[24] The parameters of these samples are listed in Table I. Prior to epitaxy of (Ga,Mn)As, we deposit buffer layers consisting of 4 nm GaAs / 30 nm $Al_{0.75}Ga_{0.25}As$ / ~ 450 nm $In_{0.15}Ga_{0.85}As$ / 30 nm GaAs (from the surface side) on a semi-insulating GaAs(001) substrate.[24-26] (Ga,Mn)As layers and top GaAs layers are grown at low temperatures between 170 and 220°C. During epitaxy, the growth front is monitored by

**TABLE I.** Channel thickness $t$, nominal Mn composition $x$, hole concentration $p_0$, carrier mobility $\mu$, and Curie temperature $T_C$ at gate electric field $E_G = 0$ for the samples investigated. The values of $p_0$ and $\mu$ are determined from the $E_G$ dependence of the sheet conductance $R_{sheet}^{-1}$ at 80 K, and $T_C$ is determined by making Arrott plots.

| Set | Device | $t$ (nm) | $x$ | $p_0$ (cm$^{-3}$) | $\mu$ (cm$^2$/Vs) | $T_C$ (K) |
|---|---|---|---|---|---|---|
| A | A1 | 3.5 | 0.065 | $0.56 \times 10^{20}$ | 1.1 | 45.7 |
|   | A2 | 4.0 |       | $0.74 \times 10^{20}$ | 2.4 | 55.1 |
|   | A3 | 4.5 |       | $2.10 \times 10^{20}$ | 3.1 | 66.8 |
|   | A4 | 5.0 |       | $2.04 \times 10^{20}$ | 3.5 | 88.0 |
| B | B1 | 4.5 | 0.052 | $1.56 \times 10^{20}$ | 4.6 | 64.4 |
|   | B2 |     | 0.072 | $1.53 \times 10^{20}$ | 3.5 | 81.7 |
|   | B3 |     | 0.089 | $4.38 \times 10^{20}$ | 3.2 | 116.9 |
| C | C1 | 4.0 | 0.075 | $3.06 \times 10^{20}$ | 3.1 | 75.5 |
|   | C2 |     | 0.100 | $5.86 \times 10^{20}$ | 2.2 | 118.2 |
|   | C3 |     | 0.125 | $7.78 \times 10^{20}$ | 1.9 | 131.7 |
|   | C4 |     | 0.175 | $11.7 \times 10^{20}$ | 1.7 | 165.3 |
|   | C5 |     | 0.200 | $15.9 \times 10^{20}$ | 1.1 | 151.7 |

*in situ* reflection high-energy electron diffraction (RHEED). For all the (Ga,Mn)As samples, RHEED shows clear streaky ×2 patterns along the [-110] direction,[5] showing two-dimensional layer-by-layer growth. Growth rate is determined to be ~12 nm/min by RHEED oscillation of GaAs at high substrate temperature of ~570°C. The Mn composition $x$ is determined from the extrapolated curve of the relationship between Mn cell temperature and $x$ for thick (> 50 nm) Ga$_{1-x}$Mn$_x$As layers in the range of $x > 0.08$, where $x$ is obtained from x-ray diffraction.[5] Because thin (Ga,Mn)As layers very often become insulating,[27,28] which prohibits probing of magnetic properties by magnetotransport measurements,[17] we consider samples with a relatively high values of Mn composition $x > 0.05$ in this work, in which the hole density is sufficiently large to preclude localization. Most of (Ga,Mn)As layers studied here have a magnetic easy axis perpendicular to the film plane due to the tensile strain introduced by the (In,Ga)As buffer layer via strain dependent magnetic anisotropy.[29,30] In samples with $x > 0.17$ the easy axis tend to be in-plane owing to a larger lattice constant of (Ga,Mn)As, resulting in a reduction in strain in the films.[24]

First, samples are annealed in air at 180°C for 1 min for set A or 5 min for sets B and C, then processed into an FET structure with a Hall-bar geometry in order to perform four-terminal measurements to determine the sheet and Hall resistances ($R_{sheet}$ and $R_{Hall}$) under gate electric fields. A Hall bar with 30-μm-wide and 150-μm-long channel is defined by photolithography and wet etching. A gate insulator with thickness $d = 40$ nm, Al$_2$O$_3$ for sets A and B or HfO$_2$ for set C, is deposited on the channel at 150°C by ALD.[22] The advantage of ALD is not only the ability to deposit high-$\kappa$ materials characterized by high coverage and good flatness but also the fact that the high-quality insulator is deposited at low temperatures, which is important for (Ga,Mn)As to avoid an unintentional annealing effect.[31] We used H$_2$O and tri-methyl-aluminum (CH$_3$)$_3$Al as precursors for Al$_2$O$_3$ and H$_2$O and tetra-kis-di-methyl-amino-hafnium Hf[N(CH$_3$)$_2$]$_4$ for HfO$_2$. The evaporation and lift-off of topmost source, drain, gate, and probe metal electrodes of 5 nm Cr / 100 nm Au complete the device structure. The values of dielectric constant $\kappa$ are 7.5 and 20 for Al$_2$O$_3$ and HfO$_2$, respectively, which are determined in a separate capacitance measurement on Au / insulator / Au capacitors.[19,22]

For the transport measurements Cu wires are bonded by In to the electrodes. The devices are mounted on a cold finger of $^4$He closed-cycle refrigerator. The Hall and sheet resistances are measured as a function of both external magnetic field $|\mu_0 H| < 0.5$ T applied in the direction perpendicular to the plane and the gate electric field $E_G$ in the range from -7 to 7 MV/cm for various temperatures in the vicinity of $T_C$.

### III. EXPERIMENTAL RESULTS

We first focus on the results obtained for the representative device with $x = 0.072$, $t = 4.5$ nm, and an Al$_2$O$_3$ insulator (device B2). The inset to Fig. 1 illustrates dependence of sheet conductance, the inverse of $R_{sheet}$, on $E_G$ at 80 K for devices A1, B2, and C5. The arrows indicate the sweep direction of $E_G$. The data point to a small hysteresis, which is presumably caused by carrier trapping by interface states between the insulator and (Ga,Mn)As and/or defects in the oxides.[32] Device C5 with a $HfO_2$ insulator shows a larger hysteresis than devices A1 and B2 with an $Al_2O_3$ insulator, suggesting a higher charge-trapping density in $HfO_2$. In order to avoid the effect of carrier trapping, we have used the slope $dR_{sheet}^{-1}/dE_G$ only in the vicinity of $E_G = 0$ to determine the carrier mobility $\mu$. Provided that the carrier mobility $\mu$ is independent of $E_G$, and that the holes are distributed uniformly over the channel thickness $t$, the sheet conductance is given by

$$R_{sheet}^{-1} = \mu e t p(E_G) = \mu e t [p_0 + \Delta p(E_G)], \quad (1)$$

where $p_0$ is hole concentration at $E_G = 0$ and $\Delta p(E_G) = p(E_G) - p_0$. For an ideal capacitor, $\Delta p$ can be determined by the capacitance of the device and the magnitude of $E_G$ as

$$\Delta p(E_G) = -\varepsilon_0 \kappa E_G /(et). \quad (2)$$

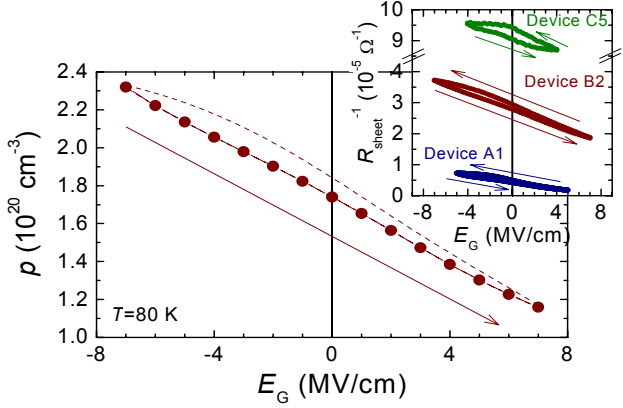

**FIG. 1.** (Color online) Hole concentration $p$ as a function of the gate electric field $E_G$ measured at $T = 80$ K for device B2. Arrows indicate the sweep direction of $E_G$. The inset shows $E_G$ dependence of sheet conductance $R_{sheet}^{-1}$ for three devices, A1, B2, and C5, where A1 and C5 have the lowest and highest conductivities, respectively, among all the devices investigated in the present work. Device C5 with $HfO_2$ gate insulator shows a larger hysteresis than A1 and B2 with $Al_2O_3$ gate insulator.

This expression for $\Delta p(E_G)$ is believed to be a good approximation for our devices as long as the swing of $E_G$ is moderate, where the hysteresis, and thus the effect of interface states is small. We determine $\mu$ as ~3.5 cm²/Vs for device B2 from the slope of the inset of Fig. 1 from

$$\mu = -[d(R_{sheet}^{-1})/d E_G]/(\kappa \varepsilon_0). \quad (3)$$

The obtained values of $\mu$ for all the devices are summarized in Table I, which are comparable to, but smaller by ~30% than, those for thick (Ga,Mn)As layers ($t$

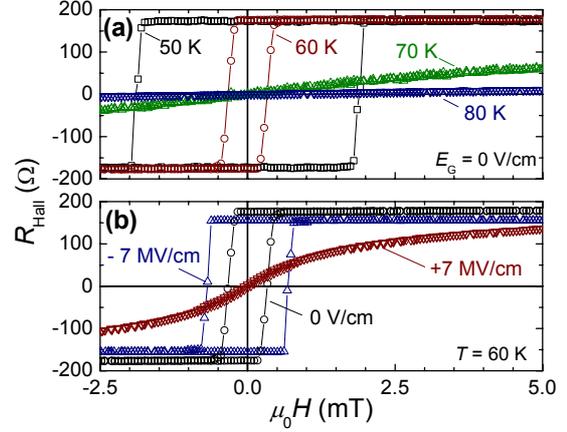

**FIG. 2.** (Color online) Magnetic field dependence of the Hall resistance $R_{Hall}$ of device B2 measured (a) at several temperatures $T = 50 - 80$ K at the gate electric field $E_G = 0$ and (b) at $T = 60$ K at $E_G = -7$, 0, and +7 MV/cm.

= 50 nm) determined from room-temperature Hall measurements.[33] The tendency of lower $\mu$ for higher $x$ is consistent with that for thick (Ga,Mn)As layers.[33]

The obtained $p(E_G) = R_{sheet}^{-1}(E_G)/(\mu et)$ is shown in the main panel of Fig. 1. This shows that the value of $p$ can be modulated twofold, from $1.2 \times 10^{20}$ to $2.4 \times 10^{20}$ cm⁻³ by changing $E_G$ from 7 to −7 MV/cm. The values of $p_0$ obtained from $R_{sheet}^{-1}$-$E_G$ curves at 80 K are summarized for all the devices in Table I. For example, the value of $p_0$ determined by room-temperature Hall measurement for device B2 is $1.23 \times 10^{20}$ cm⁻³ while $p_0$ determined from the $R_{sheet}^{-1}$-$E_G$ curve is $1.53 \times 10^{20}$ cm⁻³. This ~30% deference between $p_0$ is consistent with ~30% difference in $\mu$ determined from the two different methods, suggesting that a small but non-negligible contribution from the anomalous Hall effect affects $p_0$ value determined by Hall measurements even at room temperature.[34]

We now describe how $T_C$ varies with the gate voltage. We probe the magnetic response of (Ga,Mn)As channels by the Hall resistance $R_{Hall}$ which can be written in the form

$$R_{Hall} = (R_0/t) \mu_0 H + (R_S/t) M, \quad (4)$$

where $R_0$ and $R_S$ are the ordinary and anomalous Hall coefficients, respectively. The first term corresponds to the ordinary Hall resistance induced by the Lorentz force. The second term is the anomalous Hall resistance, which is proportional to the perpendicular component of magnetization $M$ and is dominant in (Ga,Mn)As at low temperatures.[35]

Figure 2(a) presents $R_{Hall}(H)$ at $E_G = 0$ for temperatures between 50 and 80 K. The curves at 50 and 60 K show clear square hysteresis, whereas those at 70 and 80 K show nonlinear dependence of $R_{Hall}(H)$ without hysteresis. This points to the existence of a ferromagnetic order below $T_C$ of about 80 K. The character of hysteresis loops can be dramatically changed by the application of a gate electric field. Figure 2(b) shows $R_{Hall}(H)$ for $E_G$ = −7, 0, and +7 MV/cm at $T$ = 60 K. The hystereses are more square-like under a negative $E_G$ than that under a positive $E_G$, and coercive forces are ~0.7, ~0.3, and ~ 0 mT at $E_G$ = −7, 0, and +7 MV/cm, respectively. This indicates clearly that magnetic properties can be changed by $E_G$.[17] A smaller magnitude of $R_{Hall}$ at $E_G$ = −7 MV/cm than at $E_G$ = 0 is due to the change in $R_S$, which depends on the device resistivity[36,37] modulated also by $E_G$.

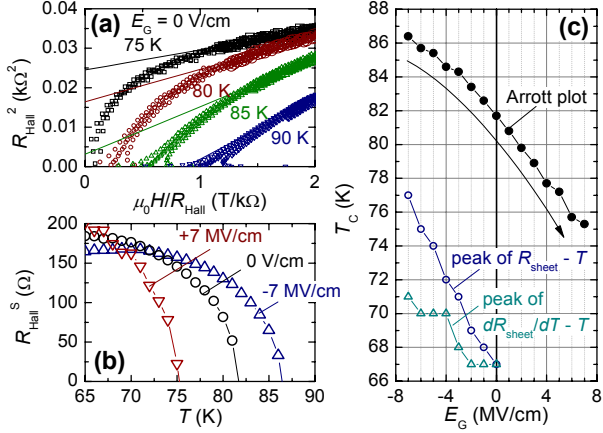

**FIG. 3.** (Color online) (a) Arrott plots for device B2 at temperatures $T$ = 75 - 90 K and at zero electric field $E_G$ = 0. (b) Temperature dependence of spontaneous Hall resistance $R_{Hall}^S$, obtained from the Arrott plots. (c) $T_C$ as function of $E_G$, as determined from the Arrott plot of $R_{Hall}$ (closed circles), the temperature dependence of $R_{sheet}$ (open circles) and the temperature dependence of $dR_{sheet}/dT$ (open triangles). The arrow indicates the sweep direction of $E_G$.

In order to see the effect of $E_G$ on $T_C$, we employ the Arrott plot of $R_{Hall}^2$ versus $\mu_0 H/R_{Hall}$ by assuming that $R_{Hall}$ is proportional to $M$.[17] Figure 3(a) shows typical Arrott plots for the data taken at $T$ = 75 – 90 K at $E_G$ = 0, where straight lines show linear fit to the data in $|\mu_0 H| > 0.25$ T. The intercept of the linear extrapolation on vertical axis corresponds to the square of the spontaneous Hall resistance $R_{Hall}^S$, which is proportional to spontaneous magnetization. The Curie temperature $T_C$ is determined as the temperature at which $R_{Hall}^S$ becomes zero. Figure 3(b) shows the $R_{Hall}^S(T)$ at $E_G$ = −7, 0, and +7 MV/cm, where a clear modulation of $T_C$ from 75 to 86 K can be seen. The dependence $T_C(E_G)$ is summarized in Fig. 3(c), where an arrow indicates the sweep direction of $E_G$. The values of $T_C$ at $E_G$ = 0 are listed in Table I for all studied samples. It is known that the temperature at which $R_{sheet}(T)$ or $dR_{sheet}/dT$ attain a maximum is one of the measure of $T_C$.[35,38] For reference, we have presented in Fig. 3(c) the values of $T_C$ determined in this way. One can see that the shift of $T_C$ induced by the gate voltage obtained from $R_{sheet}(T)$ and from $dR_{sheet}/dT$ is, respectively, slightly larger or comparable to the one determined from the Arrott plots of $R_{Hall}(T,H)$, which is another evidence of a sizable change in $T_C$ by $E_G$. Plausible origins of differences between $T_C$ values obtained from direct magnetization measurements and from critical scattering have been discussed in Ref. 38, emphasizing the role of the film thickness as well as of annealing and etching protocols. In Fig. 3(c), $T_C$ values determined from $R_{sheet}$ are presented only for negative $E_G$ because the insulating behavior of our devices makes it difficult to determine a local maximum in the temperature dependencies of $R_{sheet}$ and $dR_{sheet}/dT$ under positive $E_G$. Hereafter, we discuss the $E_G$ dependence of $T_C$ determined from the Arrott plots.

By combing the results shown in Figs. 1 and 3(c), we determine the $p$ dependence of $T_C$ as shown in Fig. 4, where a linear dependence in the logarithmic scale is observed. From the linear fit, we obtained $T_C \propto p^{0.21}$ for device B2. The same linear dependence is observed for all other devices. The results for the devices with the lowest (device A1) and the highest $p_0$ (device C5) are also shown in the main panel of Fig. 4. The inset summarizes the obtained exponent $\gamma$ in the relation $T_C \propto p^\gamma$ as a function of $p_0$ for all twelve devices. The relation of $T_C \propto p^\gamma$ with $\gamma$ = 0.19 ± 0.02 (here, an error bar is the standard deviation for all devices) is observed in a wide range of $p$ (over two decades), independent of $x$ and $t$ for the present thin (Ga,Mn)As layers, indicating that one can enhance $T_C$ by ~60% by increasing $p$ by 1 order of magnitude. The one anomalous point with $\gamma \sim 0.15$ corresponds for device A2.

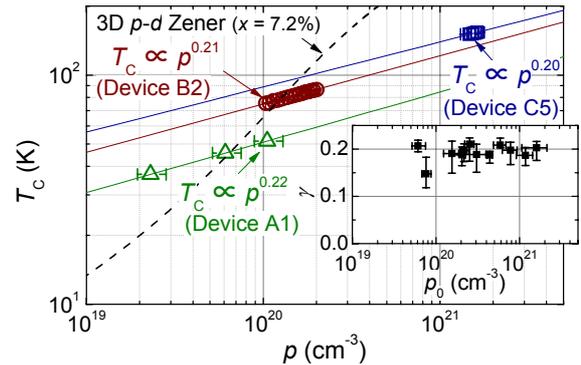

**FIG. 4.** (Color online) Logarithm plots of the Curie temperature $T_C$ versus hole concentration $p$; device B2 (circles), device A1 with the lowest $p$ (triangles), and device C5 with the highest $p$ (squares). Solid lines represent linear fit for the data. Dashed line shows the expected relation from the $p$-$d$ Zener model for the three dimensional (bulk) (Ga,Mn)As. The inset shows the exponent $\gamma$ in $T_C \propto p^\gamma$ as a function of hole concentration at $E_G$ = 0, $p_0$, for the twelve devices.

The reason of the anomaly is not clear, however, it may be related to the different quality of the insulating layer from those of the other devices. A different exponent $\gamma = 0.33$ has been obtained for thicker (Ga,Mn)As films ($t$ between 50 and 123 nm), independently of $x$ values from 0.02 and 0.085.[10,39]

## IV. THEORETICAL MODEL AND CALCULATION

The obtained relationship between $T_C$ and $p$ is not expected within the $p$-$d$ Zener model for the three-dimensional (3D) (bulk) (Ga,Mn)As,[30] where uniform carrier distribution is assumed, which predicts the exponent $\gamma = 0.6$-$0.8$ in the relevant region of hole densities, as shown by the dashed line in Fig. 4. To find the reason for this discrepancy, we note that the experimentally determined relation $T_C(p)$ has been obtained assuming that the hole concentration is uniform across the channel thickness $t$. As shown recently based on the description of magnetization measurements,[21] this assumption is not fulfilled in the relevant range of channel thicknesses for two reasons: first, according to theory of hole screening[40] and electric field distribution in $p$-$n$ junctions[41] the Thomas-Fermi screening length $\lambda$ is below 1 nm in (Ga,Mn)As, *i.e.,* much shorter than $t$. This results in a non-uniform, $z$-dependent hole modulation by the gate field $E_G$ ($z$ is along growth direction). Second, because of the existence of interface states, the hole gas is depleted near the boundary between (Ga,Mn)As and the oxide. In order to take into account the above effects, the $p$-$d$ Zener model, developed for three and two-dimensional systems,[30,42,43] was adopted to the case of thin layers of (Ga,Mn)As with a non-uniform distribution of carriers.[21] We present now the foundations of this approach.

The starting point is the observation that the hole mean free path, as determined from the mobility values, is below 1 nm (Ref. 35) and, thus, much shorter than $t$. This leads to a considerable broadening of two-dimensional subbands,[44,45] making the use of the form of the density of states suitable for three-dimensional systems more adequate for the channels in question. On the other hand, the phase coherence length $L_\phi$ of holes is much greater than $t$ (Refs. 46 and 47) so that the hole wave function is coherent across the channel thickness. This means that the spin splitting $\Delta$ of the holes is determined by spin polarization of all Mn ions residing across the channel weighted by the probability of finding a hole at a given location, $p(z)/p_S$, where the sheet hole concentration, $p_S = \int p(z) \, dz$. Under this assumption, the spin splitting $\Delta$ differs little for particular hole state and, thus, can be taken as an order parameter to be determined self-consistently in the mean-field fashion. Its magnitude is then given by[42,43]

$$\Delta = A_F \beta / (g\mu_B) \int p(z) M(z) dz / p_S, \quad (5)$$

where $A_F$ is the Fermi liquid Landau's parameter,[30,42] $\beta$ is the $p$-$d$ exchange integral, and $M(z)$ is the local magnetization of the Mn spins, brought about by a molecular field produced by spin polarization of the holes, *i.e.,* by $\Delta$ according to,

$$M(z) = g\mu_B N_0 x_{eff} S B_S[\beta S p(z) \rho_S \Delta / (4k_B T p_S)], \quad (6)$$

where $g = 2.0$ is Landé factor. $N_0 x_{eff}$ is the concentration of unpaired Mn spins in the substitutional positions assumed to be uniform across the channel,[48] and the sheet density of states (DOS) $\rho_S = \int \rho_{3D} dz$, where $\rho_{3D}$ is the three-dimensional DOS for spin excitations[30] at given hole concentration $p(z)$. By expanding the Brillouin function $B_S(y) = (S+1)y/3S$ for $y \ll 1$, we arrive to the expression for $T_C$ of a thin (Ga,Mn)As layer

$$\begin{aligned} T_C &= \frac{N_0 x_{eff} A_F \beta^2 S(S+1)}{12 k_B} \rho_S \int \frac{p(z)^2}{p_S^2} dz \\ &= \int T_C^{3D}(z) dz \cdot \int \frac{p(z)^2}{p_S^2} dz, \end{aligned} \quad (7)$$

where $S = 5/2$ is Mn spin and $T_C^{3D}$ is the Curie temperature calculated from the conventional $p$-$d$ Zener model at given hole and Mn concentration, $p(z)$ and $N_0 x_{eff}$, respectively.[6,30]

In order to determine $p(z)$ we employ the Poisson equation solver we have developed for degenerate semiconductors. In $p$-type GaAs, the hole density $p = 10^{20}$ cm$^{-3}$ corresponds to the Fermi energy $\varepsilon_F \cong 100$ meV, so that in the vicinity of $T_C$, $\varepsilon_F/k_B T$ is larger than 20 and the Fermi-Dirac rather than the Boltzmann statistics has to be employed. Because, the two-dimensional subband structure is washed out by disorder we adopt the Thomas-Fermi approximation for the determination of the charge density at a given potential profile $V(z)$. We fix the value of the effective hole mass $m^*$ at $0.9 m_0$, which is implied by the 6×6 $k \cdot p$ Hamiltonian for hole densities om the order of $10^{20}$ cm$^{-3}$.[30]

The calculations are carried out with three adjustable parameters, the net concentration of Mn acceptors $N_A$,[49] the concentration of donor-like interface states $N_i$ at insulator/(Ga,Mn)As boundary,[21,41] and the concentration of the antisite As$_{Ga}$ donors $N_D$ in the GaAs buffer adjacent to (Ga,Mn)As and grown at low temperature.[50] The other material parameters of (Ga,Mn)As, such as a dielectric constant, are assumed to be the same values as those of GaAs.[51] We considered (Ga,Mn)As without strain, since $T_C$ does not depend much on the magnitude and direction of strain.[6,30] Thus, we employ the simplified structure consisting of metal/insulator/(Ga,Mn)As/GaAs/(Al,Ga)As/GaAs substrate without (In,Ga)As buffer layer. We pin the Fermi energy at the mid gap of GaAs substrate residing 30 nm below the bottom of the (Al,Ga)As buffer layer.

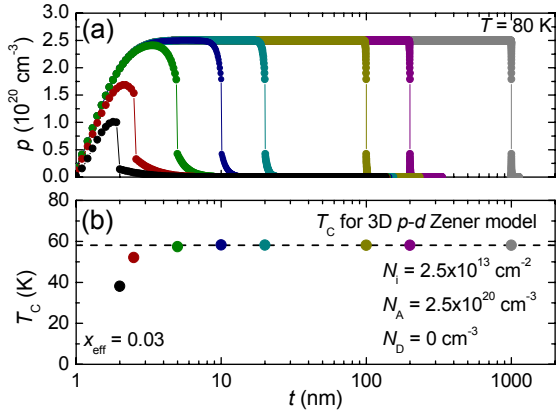

**FIG. 5.** (a) (Color online) The hole distribution profiles $p(z)$ at 80 K obtained by solving the Poisson equation for various channel thickness $t$. (b) Curie temperature $T_C$ as a function of $t$ calculated by using the determined hole distribution profiles and Eq. 7 (see main text) for Mn density $x_{eff} = 0.03$. The dotted line represents $T_C$ obtained from the $p$-$d$ Zener model for the three dimensional case (Refs. 6 and 30).

Figure 5 shows (a) the hole distribution profiles at 80 K and (b) the corresponding $T_C$ values as a function of the channel thickness for $N_A = 2.5 \times 10^{20}$ cm$^{-3}$, $N_i = 2.5 \times 10^{13}$ cm$^{-2}$, $N_D = 0$, and $x_{eff} = 0.03$ calculated as a function of the channel thickness. The dotted line in Fig. 5(b) shows $T_C^{3D}$ for the same values of $N_A$ and $x_{eff}$. Although the results depend to some extent on the magnitudes of the adjustable concentrations, generally one sees the deviation of $T_C$ from $T_C^{3D}$ in the range $t < 5$ nm, where the magnitude of $T_C$ becomes quite sensitive to the layer thickness. Actually the values of real and nominal thicknesses may differ in (Ga,Mn)As due to a formation of a native oxide layer at the film surface.[4,52] Because no quantitative information on the oxidation depth is available at this stage, we assume here the nominal thicknesses.

Now, we show how the calculation describes the experimental data for device B2. We adjust the values of $N_A$ and $N_i$ to reproduce the experimentally determined sheet hole concentration [$p_{sheet}(E_G = 0) = 6.9 \times 10^{13}$ cm$^{-2}$] at $E_G = 0$, while keeping the value of $N_D$ constant, $N_D = 4.0 \times 10^{19}$ cm$^{-3}$.[50] Figure 6 shows the computed values of (a) the energy difference between the valence band edge and the Fermi energy, $\varepsilon_V - \varepsilon_F$ and (b) the hole distribution profiles $p(z)$ at $E_G = 0$ and $T = 80$ K for various sets of $N_A$ and $N_i$, values which can reproduce the magnitude of $p_{sheet}(E_G = 0)$. The energy line-up diagram and the hole distribution profiles depend on the choice of parameters. By using the determined hole density profiles, we calculate $T_C$ by Eq. 7 for $x_{eff} = 0.03$, the outcome being summarized as a function of $N_A$ and $N_i$ in Fig. 6(c). The results indicate that the magnitude of $T_C$ scales with the value of hole concentration corresponding rather to a maximum of $p(z)$ than to $p_S$.

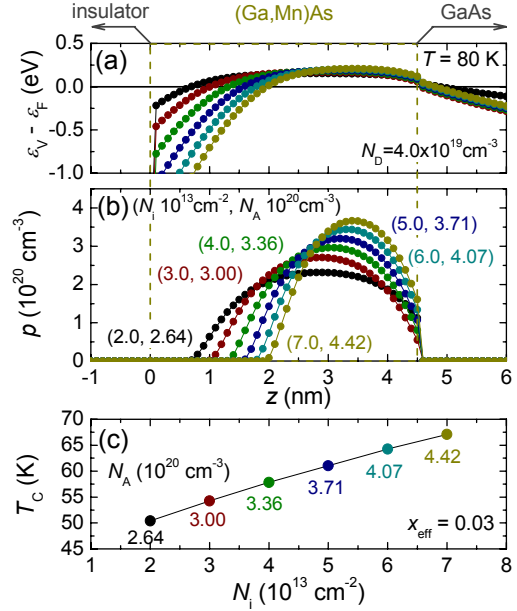

**FIG. 6.** (Color online) (a) The position of the valence band in respect to the Fermi energy $\varepsilon_V - \varepsilon_F$ and (b) hole distribution profiles $p(z)$ for various $N_i$ and $N_A$ at 80 K and for $N_D = 4.0 \times 10^{19}$ cm$^{-3}$. Thickness of (Ga,Mn)As is $t = 4.5$ nm. (c) $T_C$ calculated by using Eq. 7 with $x_{eff} = 0.03$ are shown as a function of $N_i$ and $N_A$.

We have also computed hole distribution profiles and the corresponding magnitudes of $T_C$ as a function of the gate electric field $E_G$. The outcome shows that a narrower hole distribution results in a steeper change in $T_C$ under the application of the gate voltage and that there indeed exists a set of parameters allowing one to describe the experimentally obtained relation, $T_C \propto p^\gamma$. For the device B2, the calculation with $N_A = 3.4 \times 10^{20}$ cm$^{-3}$, $N_i = 4.0 \times 10^{13}$ cm$^{-2}$, and $N_D = 4.0 \times 10^{19}$ cm$^{-3}$ produces $\gamma = 0.23$, which compares favorably with the experimentally observed $\gamma = 0.21$. The inset to Fig. 7 presents the calculated hole distribution profiles for various $E_G$ using the above parameter set, which shows a non-uniform hole modulation along $z$. The main panel in Fig. 7 depicts the dependence of $T_C$ on the normalized hole concentration $p_S/t$, where $p_S/t$ corresponds to the hole concentrations $p$, determined experimentally by using Eqs. 1--3. The measured $T_C$ values presented by circles and the calculated ones by stars are in good agreement. The solid lines show the linear fits. The effective Mn composition $x_{eff} = 0.042$ obtained by fitting the absolute magnitudes of $T_C$ is reasonable for the sample with the nominal Mn concentration $x = 0.072$.[11,53] The results indicate that our experimentally obtained relationship $T_C \propto p^\gamma$ with $\gamma = 0.19 \pm 0.02$ can be reproduced by the modified $p$–$d$ Zener model for thin films.

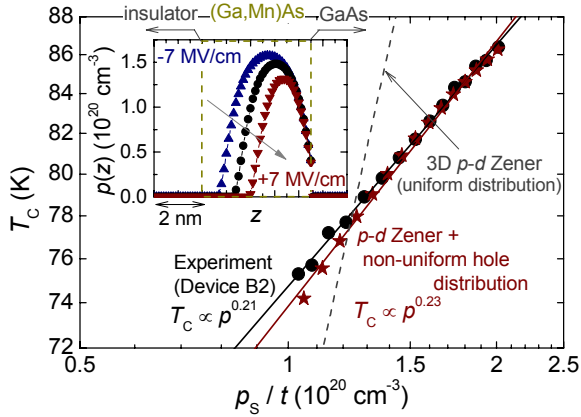

**FIG. 7.** (Color online) Double logarithmic plot showing the relation between the Curie temperature $T_C$ and the sheet hole concentration normalized by nominal thickness, $p_S/t$. Circles and stars are the values of $T_C$ measured for device B2 (same as those in Fig. 4) and calculated by Eq. 7, respectively. The solid lines show linear fits and the dashed line is the prediction of the $p$–$d$ Zener model for the three dimensional (bulk) (Ga,Mn)As. The inset presents the computed hole distribution profiles for various gate electric fields $E_G$.

So far, we have used the nominal channel thickness $t$ of (Ga,Mn)As. Even if we consider the possible existence of a native oxide-layer, whose thickness can be on the order of ~ 1 nm,[4,52] we can find a set of reasonable parameters to reproduce the experimental data, *e.g.*, $N_A = 3.8 \times 10^{20}$ cm$^{-3}$, $N_i = 3.0 \times 10^{13}$ cm$^{-2}$, and $N_D = 3.0 \times 10^{19}$ cm$^{-3}$ for the case with $t = 3.5$ nm (considering 1.0-nm thick oxide layer). Because the interface states and resulting hole distribution affect significantly the dependence of $T_C$ on the gate electric field, the reduction in interface states at the insulator / (Ga,Mn)As boundary is important to have a better control over magnetic properties by the application of electric fields, particularly if one wants to probe the effect of carrier accumulation.

## V. SUMMARY

In summary, we have examined experimentally a relationship between the Curie temperature $T_C$ and the hole concentration $p$ in thin (Ga,Mn)As films by using the field effect in metal-insulator-semiconductor structures. The relation $T_C \propto p^{0.19 \pm 0.02}$ is observed for a wide range of hole concentrations (from $10^{19}$ to $10^{21}$ cm$^{-3}$ as shown in Fig. 4) as well as for many samples with differing $x$ (from 0.052 to 0.20) and thickness $t$ (from 3.5 to 5 nm). We have employed a model for the evaluation of $T_C$, which generalizes the mean-field $p$-$d$ Zener model for the case of a nonuniform hole distribution in gated channels, determined here by solving the Poisson equation. This approach shows that the observed relation $T_C \propto p^{0.19 \pm 0.02}$ can be reproduced by adopting a reasonable set of parameters. The results are important for further understanding of the difference in magnetic and electrical properties of thin (Ga,Mn)As films from those of thick (Ga,Mn)As as well as their unique behaviors under gate electric fields.[20]

## ACKNOWLEDGEMENTS

The authors thank K. Ohtani, Y. Norifusa, T. Endoh, Y. Chiba, and M. Sakuraba as well as A. Korbecka and J. M. Majewski for useful discussions. This work was supported in part by Grant-in-Aids from MEXT/JSPS, the GCOE Program at Tohoku University. Y. N. acknowledges support from JSPS (Research Fellowship for Young Scientists). M. S. and T. D. acknowledge support from the European Research Council within "Ideas" 7th Framework Programme of European Commission (FunDMS Advanced Grant)

---